\documentclass[journal=aaemcq,manuscript=article]{achemso}
\setkeys{acs}{articletitle=true,maxauthors=0,etalmode=truncate}
\SectionNumbersOn
\usepackage{amsmath,amssymb,bm}
\usepackage{graphicx}
\usepackage{booktabs}
\usepackage{siunitx}
\usepackage{microtype}
\usepackage{url}
\usepackage{xcolor}

\newcommand{\LiPDOS}{Li-PDOS}
\newcommand{\Wone}{W_1}
\title{Lithium-Projected Phonon Spectral Distributions as Robust Descriptors of Ionic Conductivity in Solid Electrolytes}
\author{Gajendra Bohara}
\affiliation{Department of Physics and Astronomy, Clemson University, Clemson, South Carolina 29634, United States}
\author{Ramakrishna Podila}
\email{rpodilar@g.clemson.edu}
\affiliation{Department of Physics and Astronomy, Clemson University, Clemson, South Carolina 29634, United States}

\keywords{solid-state electrolyte, ionic conductivity, phonon density of states, lithium-projected phonons, machine-learning interatomic potential, Wasserstein kernel, materials informatics}
\begin{document}
\begin{abstract}
Lattice dynamics are widely invoked in the design of solid electrolytes, yet phonon information is commonly compressed into a band center or another scalar softness measure. Here we test whether the complete lithium-projected phonon density of states (\LiPDOS) provides a reproducible descriptor of experimental room-temperature ionic conductivity. MatterSim forces and Phonopy were used to generate harmonic total and Li-projected spectra for crystallographically resolved entries in the OBELiX dataset. A composition and structure audit defined a primary cohort of 260 materials (212 train and 48 test), a strict cohort of 241, and an exact-composition cohort of 168. Across 20 independently generated phonon-calculation database comparisons, the mean Wasserstein-1 distance was 0.542 THz for total DOS and 0.731 THz for \LiPDOS, revealing broad agreement but systematic, projection-dependent softening. Higher conductivity was associated with redistribution of normalized Li spectral weight toward low frequencies: in the untouched test set, the Li fractions below 2 and 5 THz had Spearman coefficients of 0.333 and 0.374, while the 5\% cumulative-frequency quantile had a coefficient of $-0.393$. A Wasserstein kernel on the full Li-PDOS achieved held-out $R^2=0.444$, compared with 0.012 for total DOS and 0.181 for a static composition--structure kernel. The Li model remained stable in the strict ($R^2=0.462$) and exact ($R^2=0.451$) cohorts. Family adjustment attenuated scalar associations, and Li-versus-total whole-spectrum dependence was cohort sensitive. The results therefore support mobile-ion-resolved spectral distributions as useful comparative screening descriptors, not as a universal causal softness law.
\end{abstract}

\section{Introduction}

All-solid-state batteries are being pursued because inorganic solid electrolytes could enable nonflammable cells, lithium-metal anodes, and higher energy densities than conventional liquid-electrolyte architectures.\cite{manthiram2017,janek2016,albertus2018,randau2020} Their practical realization, however, requires more than high ionic conductivity. A viable solid electrolyte must also exhibit negligible electronic conductivity, chemical and electrochemical compatibility, mechanical integrity, and manufacturability.\cite{bachman2016,famprikis2019,kerman2017,banerjee2020,xiao2020,zhao2020} These requirements are strongly coupled: compositional substitutions that modify carrier concentration or migration barriers can also alter phase stability, elastic response, and interfacial behavior, while current localization and mechanical defects may promote lithium penetration even through nominally stable ceramic electrolytes.\cite{porz2017,han2019}

Ionic transport in a crystalline solid is therefore not determined by equilibrium geometry alone. The measured conductivity reflects carrier concentration, defect thermodynamics, migration-network connectivity, local coordination, correlated hopping, and the free-energy response of the surrounding lattice.\cite{wang2015,bachman2016,famprikis2019,he2020} Lattice dynamics can influence this response through several complementary mechanisms. Low-frequency collective distortions may transiently widen migration bottlenecks, particular eigenvectors may align with ion-hopping coordinates, rotational motion of complex anions may couple to mobile ions, and anharmonic fluctuations may reshape the migration free-energy landscape.\cite{li2015,muy2018,krauskopf2017,krauskopf2018,muy2021,gordiz2021,xu2022,song2024,ouyang2024,ding2025,pham2026} Mode-selective excitation and time-resolved measurements further suggest that a limited subset of vibrations can affect transport disproportionately.\cite{gordiz2021,pham2026}

These observations have motivated the use of lattice-dynamical descriptors in solid-electrolyte screening. Li-phonon band centers and related softness measures have been connected to migration energetics and used in high-throughput searches for candidate conductors.\cite{muy2018,muy2019} Machine-learning (ML) approaches have likewise progressed from static composition--structure descriptors to automated structural representations and phonon-informed features.\cite{sendek2019,lu2022,jaafreh2024,kim2024} More recently, high-throughput calculations have shown that phonon-softness descriptors can enrich lithium-superionic-conductor (LISICON) candidates and guide molecular-dynamics validation.\cite{aghoghovbia2026} At the same time, cross-chemistry studies have reported weak or inconsistent relationships between mean phonon frequencies and migration barriers, particularly when the contributing eigenvectors are poorly aligned with the relevant diffusion pathway.\cite{sagotra2019,greene2024,ouyang2024} Taken together, these studies establish the value of lattice dynamics while also exposing a limitation of most existing approaches: the vibrational spectrum is typically compressed into a band center, Debye frequency, selected peak, overlap integral, or another engineered scalar.

The central question addressed here is whether the \emph{complete mobile-ion-projected spectrum} contains transferable information about experimentally measured ionic conductivity that is not preserved by such scalar summaries. This question requires distinguishing both the physical observable and its mathematical representation. Total density of states (DOS) combines mobile-ion and framework vibrations, so a Li-specific transport signature may be diluted by modes dominated by the host sublattice. Nevertheless, framework motion and Li--anion coupling remain integral to transport, and the projected spectrum should not be interpreted as an isolated-ion quantity.\cite{gordiz2021,xu2022,ouyang2024,pham2026} In addition, after unit-area normalization, a phonon spectrum is a nonnegative distribution over an ordered frequency axis. Treating frequency bins as independent Cartesian coordinates ignores the distinction between shifting spectral weight over nearby and widely separated frequencies and introduces constant-sum constraints familiar from compositional data analysis.\cite{aitchison1982} Cumulative-distribution and Wasserstein representations instead retain the ordering and metric structure of the frequency coordinate and provide suitable kernels for distribution-valued inputs.\cite{kolouri2016,bachoc2018}

We investigate these questions using the experimentally grounded OBELiX dataset, which provides curated room-temperature ionic conductivities and a fixed, leakage-aware train/test split across chemically diverse solid electrolytes.\cite{therrien2026} Harmonic vibrational spectra are generated at scale using the MatterSim universal machine-learning interatomic potential (MLIP) and PhonoPy, enabling a common computational treatment of materials that would be prohibitively expensive to evaluate individually using first-principles calculations.\cite{yang2024} For every material, we analyze both the total DOS and the lithium-projected DOS (Li-PDOS), and compare scalar descriptors with representations of the complete spectral distribution.

The principal innovation is the identification of an experimentally anchored, mobile-ion-resolved spectral redistribution associated with ionic transport. Higher-conductivity materials exhibit not merely a reduction in a characteristic Li frequency, but a reproducible transfer of normalized Li spectral weight toward the low-frequency region, accompanied by compensating changes at intermediate and higher frequencies. This pattern is identified using the training cohort and reproduced directionally in the untouched OBELiX test set. Representing the complete Li-PDOS through Wasserstein geometry captures this redistribution directly and yields substantially stronger held-out prediction than either the corresponding total-DOS representation or the tested static composition--structure baseline. The performance of the Li-distribution model also remains stable across progressively restrictive composition-audit cohorts. These results show that the frequency-resolved distribution of mobile-ion vibrational weight contains transferable information about experimental conductivity that is not retained by a single softness parameter or by the total phonon spectrum.

To establish that this result does not arise from a particular structural-processing choice, frequency grid, censoring convention, or unstable phonon calculation, we use a validation-first analysis. Crystal compositions are audited before modeling, including explicit identification of finite-cell ordering approximations. Calculated spectra are compared with exact-phase reference calculations from the National Institute for Materials Science (NIMS) Materials Data Repository phonon calculation database (PhononDB),\cite{togoNIMSPhononDB,togo2023phonopy} and numerical-convergence, imaginary-mode, frequency-warp, conductivity-censoring, and audit-cohort sensitivities are evaluated. Frequency-localized associations are selected exclusively from the training data before examination in the untouched test set, and all comparisons between static, total-DOS, and Li-PDOS representations are performed on identical material cohorts.

These safeguards also clarify the scope of the discovery. Universal MLIPs can exhibit systematic phonon softening and chemistry-dependent force errors,\cite{loew2025,deng2025,du2025} and several scalar associations weaken after accounting for electrolyte family. Moreover, the relative whole-spectrum dependence of total DOS and Li-PDOS varies in the smallest exact-composition cohort. These observations limit a universal mechanistic interpretation of any single harmonic descriptor, but they do not alter the main result: mobile-ion-resolved spectral redistribution provides a reproducible and quantitatively useful description of experimental ionic-conductivity variation across chemically diverse solid electrolytes.

\section{Experimental Section}
\subsection{OBELiX Targets, Splits, and Audit Cohorts}
The experimental targets and tabulated crystallographic metadata were taken from OBELiX.\cite{therrien2026} The published dataset contains 599 entries, comprising 478 training and 121 test records. We preserved the official split throughout. Conductivity values were converted to
\begin{equation}
 y=\log_{10}\!\left[\frac{\sigma}{1\ \mathrm{S\,cm^{-1}}}\right].
\end{equation}
Entries reported as upper limits were assigned the reported limit in the primary analysis; half-limit substitution and exclusion were evaluated as sensitivities.

Paired total-DOS and Li-PDOS files were calculated for 311 entries where crystal information files (cif) was available. The production structure associated with each spectrum was compared with the intended reduced composition and, where available, the pre-relaxation source CIF. Composition matching was invariant to the number of formula units. Because finite ordered approximants of partially occupied structures require integer atom counts, the audit distinguished exact proportional matches (\texttt{PASS}), acceptable nearest-integer realizations (\texttt{PASS\_INTEGERIZED\_COMPOSITION}), and limited finite-cell approximations (\texttt{WARN\_INTEGERIZATION\_APPROXIMATION}). Element-set errors, general stoichiometry failures, Li-vacancy filling errors, severe atomic overlap, unreadable structures, and absent audit records were excluded. A detailed list is provided in the supporting information. 

The primary cohort retained all three accepted audit categories and contained 212 training and 48 test materials. The strict cohort excluded the warning-level approximations (198/43), and the exact cohort retained only proportional matches (133/35). These nested cohorts were used to test whether the scientific conclusions depended on accepting integerized or approximate representations of disordered materials. Family labels were normalized before within-family permutations and leave-one-family-out analyses.

\subsection{MatterSim--Phonopy Workflow}
Crystal structures were parsed with pymatgen,\cite{ong2013} and disordered sites were converted to an adaptive ordered approximation by sampling integer supercells while minimizing compositional error. Ordered structures were perturbed by a 0.03 \AA{} random displacement and relaxed with the MatterSim calculator on a GPU.\cite{yang2024} A FIRE pre-relaxation used a force threshold of $10^{-3}$ eV \AA$^{-1}$ and at most 500 steps, followed by variable-cell Broyden–Fletcher–Goldfarb–Shanno (BFGS) optimization to $10^{-5}$ eV \AA$^{-1}$ with at most 1500 steps. Atomic operations were implemented through the Atomic Simulation Environment.\cite{larsen2017}

Harmonic force constants were generated by finite displacement using Phonopy.\cite{parlinski1997,togo2023jpsj,togo2023jpcm} The production configuration used a $2\times2\times2$ supercell, displacement amplitude 0.03 \AA, symmetry tolerance $10^{-3}$, force-constant symmetrization, and a $20\times20\times20$ reciprocal-space mesh with eigenvectors retained. Total DOS was obtained from the mesh, while Li projections were summed over Li-site eigenvector contributions.  Universal MLIPs are now accurate enough for broad phonon screening, but systematic softening remains possible and motivated the validation and cohort-sensitivity analyses below.\cite{loew2025,deng2025,du2025}

\subsection{Spectrum Harmonization and Descriptors}
Only finite frequencies satisfying $0<\nu\leq100$ THz were included. Small negative DOS values caused by numerical broadening were clipped to zero. Total DOS and Li-PDOS were interpolated to common bins and normalized independently,
\begin{equation}
 p(\nu)=\frac{g(\nu)}{\int_0^{100\,\mathrm{THz}}g(\nu')\,d\nu'}.
\end{equation}
The primary association grid used 1 THz bins, with 0.5 and 2 THz sensitivity analyses. Raw integrated areas were retained only for quality control because they depend on cell size and projection normalization.

Prespecified scalar descriptors included spectral fractions below 2 and 5 THz, centroid frequency, and the frequency $q_{05}$ containing the first 5\% of cumulative weight. Additional moments, quantiles, entropy, participation, and fixed-window features are evaluated in the Supporting Information. Quartile-averaged spectra were defined from training targets only. The high-minus-low difference curve and its 95\% bootstrap interval were computed by resampling materials within the highest and lowest training quartiles.

\subsection{External Reference Validation and Numerical Convergence}
Potential reference phases were screened by composition, space group, and pymatgen structure matching. Only strict phase matches entered aggregate validation. Twenty comparisons were retained, corresponding to total DOS and Li-PDOS for ten OBELiX records spanning Li$_3$PO$_4$, Li$_3$ClO, Li$_3$OBr, Li$_3$AlF$_6$, Li$_3$N, LiZnPS$_4$, Li$_3$SbS$_4$, and LiGe$_2$(PO$_4$)$_3$. Reference force constants and structures were obtained from NIMS/PhononDB records linked to Materials Project identifiers.\cite{jain2013,togo2023jpsj,togo2023jpcm} Each model and reference spectrum was normalized over its common positive-frequency interval. We report the one-dimensional Wasserstein distance
\begin{equation}
\Wone(p,q)=\int_{0}^{\infty}
\left|F_p(\nu)-F_q(\nu)\right|,d\nu,
\end{equation}
where
\begin{equation}
F_p(\nu)=\int_{0}^{\nu}p(\nu'),d\nu',
\qquad
F_q(\nu)=\int_{0}^{\nu}q(\nu'),d\nu'
\end{equation}
are the cumulative distribution functions of the unit-area-normalized spectra (p($\nu$)) and (q($\nu$)), respectively. We also report the spectral overlap $O=\int_{0}^{\infty}\min\left[p(\nu),q(\nu)\right],d\nu$
together with the centroid error, low-frequency-fraction error, and ($q_{05}$) error.

Convergence tests varied displacement amplitude, reciprocal-space mesh, and supercell for five representative structures. Changes in full spectral distance and in the low-frequency descriptors were evaluated relative to the production configuration. These calculations were used to bound numerical sensitivity, not to imply that a numerically converged MLIP spectrum is identical to density functional theory (DFT).

\subsection{Association Tests and Whole-Spectrum Dependence}
At each frequency bin, Pearson and Spearman associations with $y$ were computed. Bootstrap intervals used 3000 resamples in the final cohort-sensitivity. Training-set multiplicity was controlled with a 5000-permutation maximum-absolute-statistic procedure and Benjamini--Hochberg false-discovery-rate adjustment.\cite{efron1979,benjamini1995} Mutual information, distance correlation, and spline gains were used as complementary nonlinear diagnostics.\cite{kraskov2004,szekely2007} The test split was not used to choose frequency windows.

Whole-spectrum dependence was quantified with normalized Hilbert--Schmidt independence criterion (HSIC).\cite{gretton2005} Bandwidths were set by unsupervised median distances. Significance was evaluated using 5000 global permutations and 5000 permutations constrained within normalized electrolyte-family labels. A separate analysis tested each spectral kernel against out-of-fold residuals from the static model. Direct Li-versus-total comparisons used paired label permutations.

\subsection{Distribution-Aware Prediction}
Static features included elemental fractions, composition-weighted elemental statistics, space-group number, and lattice parameters. Missing values were imputed within the fitting pipeline and standardized. The static baseline used an radial basis function (RBF) kernel. Spectral models included ordinary vector RBF, centered-log-ratio RBF, cumulative-distribution RBF, and Wasserstein kernels. The one-dimensional Wasserstein representation was implemented through quantile embeddings, retaining the ordering of the frequency coordinate.\cite{aitchison1982,kolouri2016,bachoc2018}

Kernel ridge regression used grouped nested cross-validation on the official training set. Hyperparameters were selected without access to the official test labels. The final model was fit to the complete training cohort and evaluated once on the corresponding official test subset. Paired differences were estimated from 3000 bootstrap samples. Strict and exact sensitivity models reused the primary-training hyperparameters, so changes reflect cohort definition rather than repeated target-driven tuning. Ridge, tree-based, scalar-only, fusion, stacking, and formal censored-regression alternatives are reported in the Supporting Information.\cite{hoerl1970,chen2016}

\section{Results and Discussion}
\subsection{Audit-First Cohort Construction}
Figure S1 summarizes data availability and cohort construction. Of 599 OBELiX records, valid CIF was used to produce 311  paired spectra. The audit retained 260 in the primary cohort, 241 in the strict cohort, and 168 in the exact-composition cohort. The reduction is scientifically important: crystallographic disorder, integerization, and vacancy preservation are not bookkeeping details when species-projected spectra are interpreted. A structure with the correct framework but an incorrectly filled Li sublattice can alter both the number and character of mobile-ion modes.

The final cohort spans garnets, argyrodites, LISICON-related materials, oxides, halides, phosphates, sulfides, nitrides, antiperovskites, and smaller families. Coverage is highly imbalanced, and the test conductivity distribution remains broad, extending over more than ten orders of magnitude. Consequently, aggregate performance can be dominated by cross-family differences, while per-family estimates can have large uncertainty. This motivates the nested audit cohorts and family-constrained tests rather than a single unqualified pooled analysis.

\subsection{Independent DFT References Bound the Interpretation of MLIP Spectra}
\begin{figure*}[t]
\centering
\includegraphics[width=\textwidth]{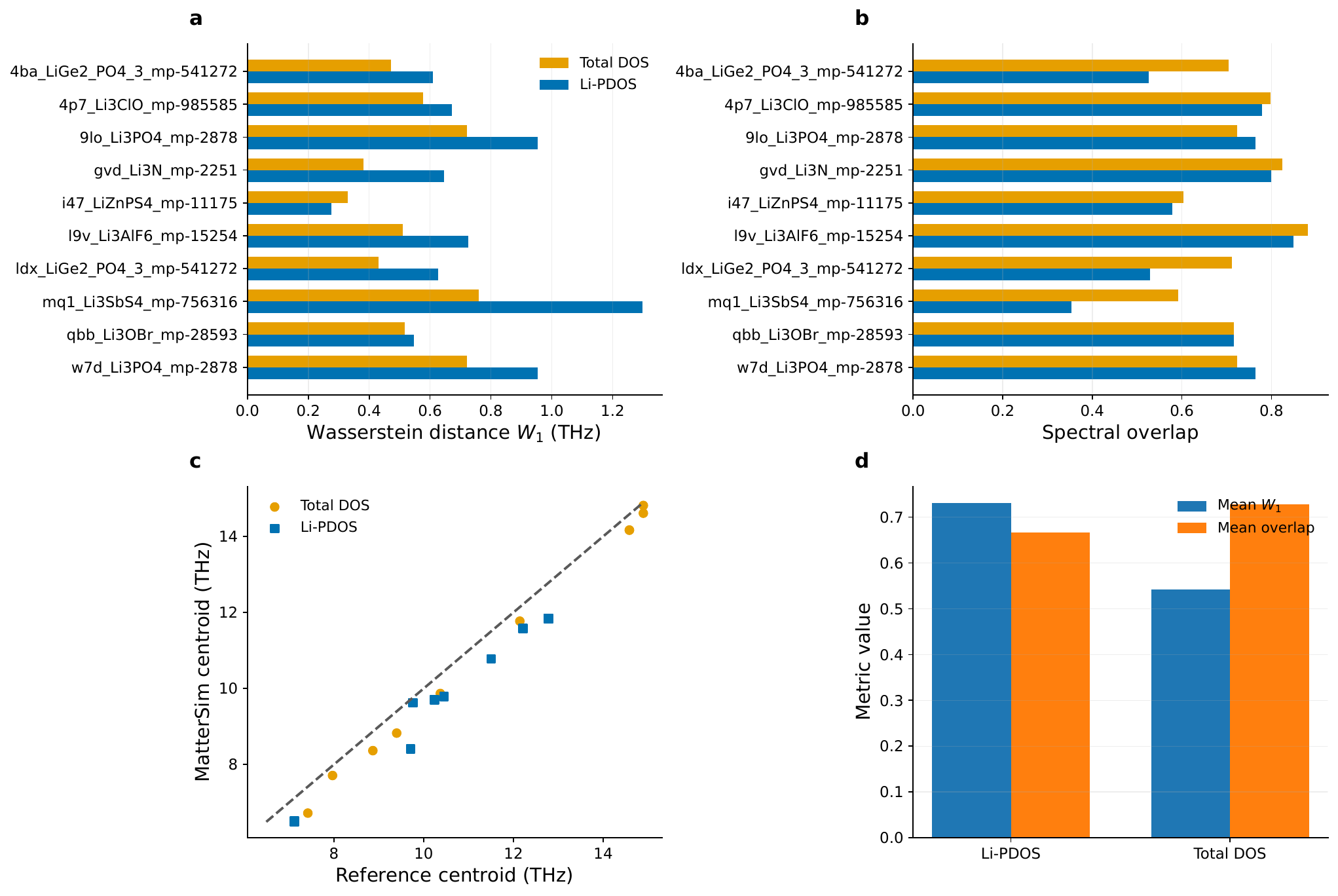}
\caption{External validation against strict-phase NIMS/PhononDB references. (a) Wasserstein-1 distances for total DOS and Li-PDOS in each accepted OBELiX--reference comparison. (b) Spectral overlaps. (c) Model versus reference centroids with the identity line. (d) Aggregate validation statistics for the two representations. Each spectrum was independently normalized over the common positive-frequency interval. Complete spectral overlays and per-case plotted data are provided in Figs S2-S21.}
\label{fig:validation}
\end{figure*}

The reference comparison in Figure~\ref{fig:validation} shows that MatterSim--Phonopy captures broad frequency ranges and cumulative shapes but does not reproduce every projected feature quantitatively. Across ten accepted material cases, mean total-DOS $\Wone$ was 0.542 THz, median $\Wone$ was 0.513 THz, and mean overlap was 0.728. The corresponding Li-PDOS values were 0.731 THz, 0.659 THz, and 0.666. Mean centroid errors were $-0.410$ THz for total DOS and $-0.711$ THz for \LiPDOS. Thus, the universal potential is systematically softer on average, and the discrepancy is larger for the mobile-ion projection. Detailed external reference overlays are presented in the Supporting Information Figs. S2-S21. 

The case-by-case structure matters more than the average. LiZnPS$_4$ shows a relatively small Li-projected distance ($\Wone=0.275$ THz), whereas Li$_3$SbS$_4$ has $\Wone=1.298$ THz and lower overlap. Two independently processed Li$_3$PO$_4$ records yield nearly identical errors, suggesting that their shift is associated with the model/reference combination rather than duplicate-entry variability. The comparison also demonstrates why total-DOS agreement cannot validate a species projection: force errors localized to the Li environment can be partly hidden by framework-dominated modes.

The validation supports statistical analysis of broad spectral distributions and cumulative descriptors, but it argues against assigning mechanistic significance to an individual calculated peak without mode-level or finite-temperature corroboration. The observed softening is consistent with broader assessments of universal MLIP phonons.\cite{loew2025,deng2025} Numerical convergence further indicates that the selected low-frequency descriptors are generally more stable than the full spectral distance, although sensitivity remains material dependent (Figs. S2-S21). We therefore require claims to survive aggregation over neighboring frequencies, held-out replication, and audit/preprocessing sensitivities.

\subsection{Conductivity Is Associated with Spectral Redistribution, Not a Single Softness Number}
\begin{figure*}[t]
\centering
\includegraphics[width=\textwidth]{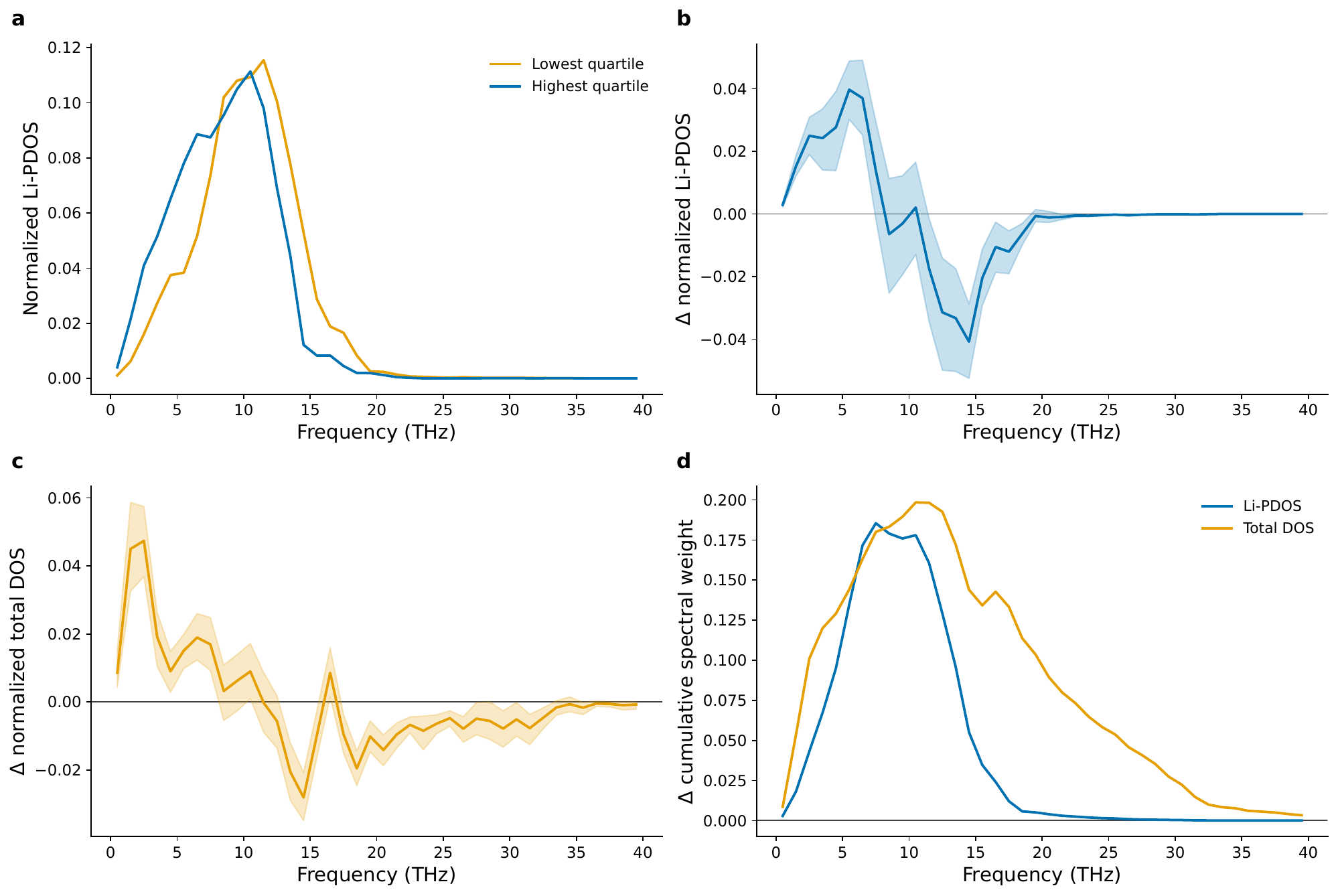}
\caption{Conductivity-stratified spectral redistribution in the primary training cohort. (a) Mean normalized \LiPDOS~for the lowest and highest conductivity quartiles. (b) Highest-minus-lowest-quartile \LiPDOS; shading is the 95\% bootstrap interval. (c) Corresponding high-minus-low difference for total DOS. (d) Cumulative high-minus-low spectral-weight differences for \LiPDOS~and total DOS. Quartile thresholds were defined only from training targets.}
\label{fig:redistribution}
\end{figure*}

Training materials in the highest conductivity quartile have more normalized Li weight at low frequency than those in the lowest quartile (Figure~\ref{fig:redistribution}a). The high-minus-low curve is positive primarily below approximately 8 THz and negative over parts of the 10--18 THz region (Figure~\ref{fig:redistribution}b). Total DOS shows a related but broader redistribution, with positive low-frequency weight and compensating depletion extending to higher frequencies (Figure~\ref{fig:redistribution}c). Because each spectrum has unit area, ``more low-frequency weight'' necessarily denotes a redistribution rather than an unconstrained increase in vibrational states. The cumulative difference (Figure~\ref{fig:redistribution}d) emphasizes that Li and total spectra redistribute over different frequency ranges.

The physical interpretation must remain conditional. Earlier family-specific studies connected low band centers to lower migration barriers,\cite{muy2018,krauskopf2017,krauskopf2018} while mode-resolved and anharmonic studies demonstrate that alignment, local bonding, and coupling can matter more than global softness.\cite{gordiz2021,xu2022,ouyang2024,greene2024,pham2026} Our pooled redistribution is compatible with weaker effective Li restoring forces or greater access to collective distortions, but it does not establish which normal modes participate in a hop, nor does a harmonic DOS encode finite-temperature defect populations.

\subsection{Frequency-Resolved Effects Replicate Directionally in the Official Test Set}
\begin{figure*}[t]
\centering
\includegraphics[width=\textwidth]{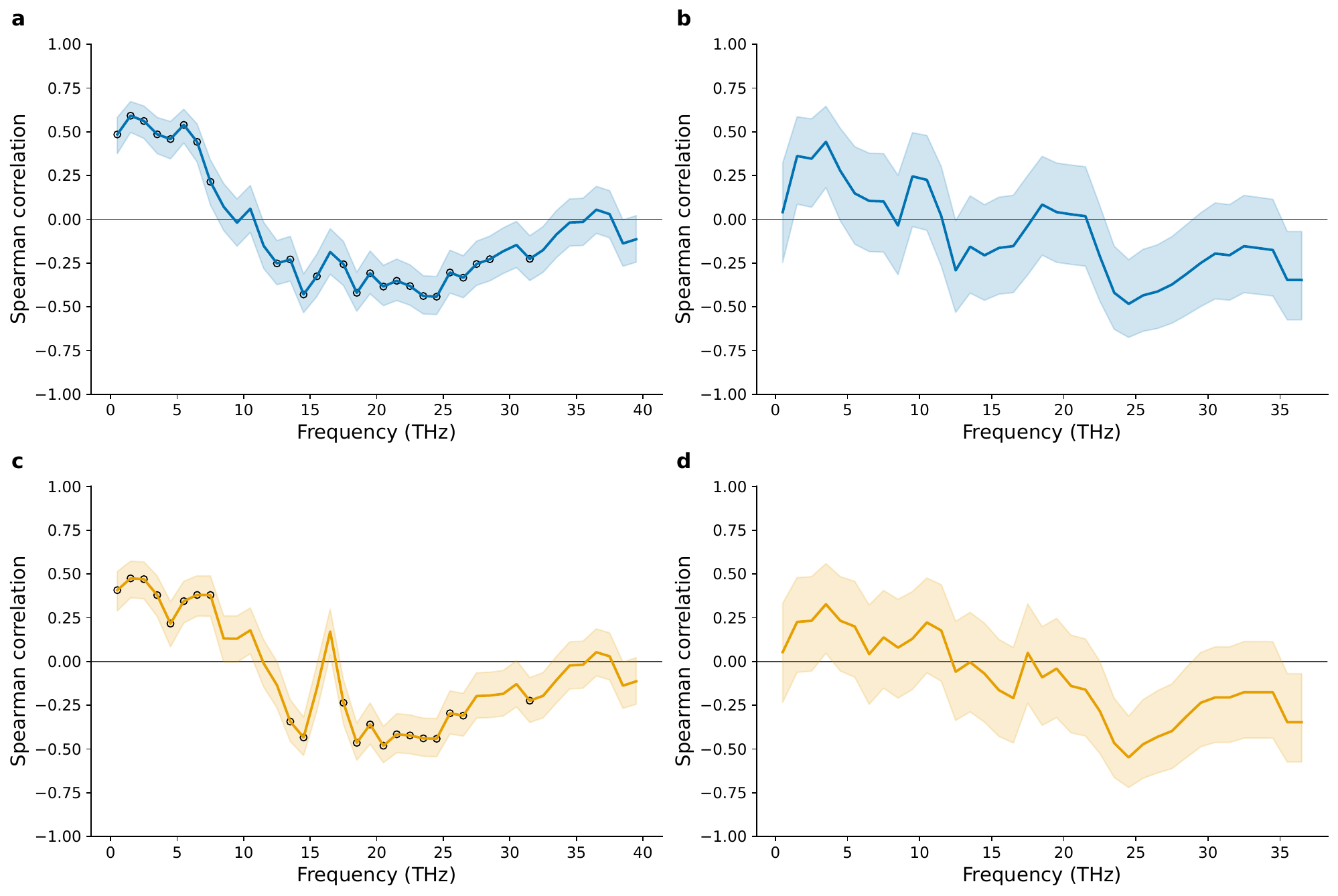}
\caption{Frequency-resolved Spearman associations with experimental log conductivity. (a,b) \LiPDOS in the official training and test cohorts. (c,d) Total DOS in training and test. Shading denotes 95\% intervals. Open circles in training indicate bins passing the 5000-permutation maximum-statistic correction. The test split was not used to select windows or tune models.}
\label{fig:frequency}
\end{figure*}

The training scan localizes the strongest positive Li-PDOS associations below approximately 7 THz and negative associations over broad intermediate-frequency ranges (Figure~\ref{fig:frequency}a). Open markers identify training bins that survive the maximum-statistic correction. The untouched test set reproduces a positive low-frequency region centered near 1--4 THz and a negative region near approximately 23--28 THz, although its confidence bands are wider because $n=48$ (Figure~\ref{fig:frequency}b). Total DOS shows a similar low-frequency direction but a more pronounced negative intermediate-frequency region (Figure~\ref{fig:frequency}c,d).

Prespecified cumulative descriptors provide a less multiplicity-sensitive summary. In the primary test cohort, the Li fractions below 2 and 5 THz have Spearman coefficients $\rho=0.333$ (95\% CI 0.038--0.588) and $\rho=0.374$ (0.079--0.616), respectively. The Li $q_{05}$ is negatively associated with conductivity, $\rho=-0.393$ ($-0.644$ to $-0.089$), indicating that the cumulative Li spectrum begins earlier in more conductive materials. These directions persist in the strict test cohort and strengthen in the exact test cohort: the exact-cohort values are 0.552, 0.556, and $-0.545$, respectively. Total-DOS centroid also replicates negatively in the primary test set ($\rho=-0.429$), underscoring that framework dynamics retain information even when the Li projection gives the strongest predictive model.

The train/test design is essential. The frequency scan is exploratory within training after multiplicity correction; the official test curves are a directional replication rather than a second feature-selection exercise. Sensitivity to 0.5 and 2 THz bins, frequency warping, censoring policy, and imaginary-mode filtering is shown in Figures S22-S25.

\subsection{Family Structure Attenuates Scalar Associations}
\begin{figure*}[t]
\centering
\includegraphics[width=\textwidth]{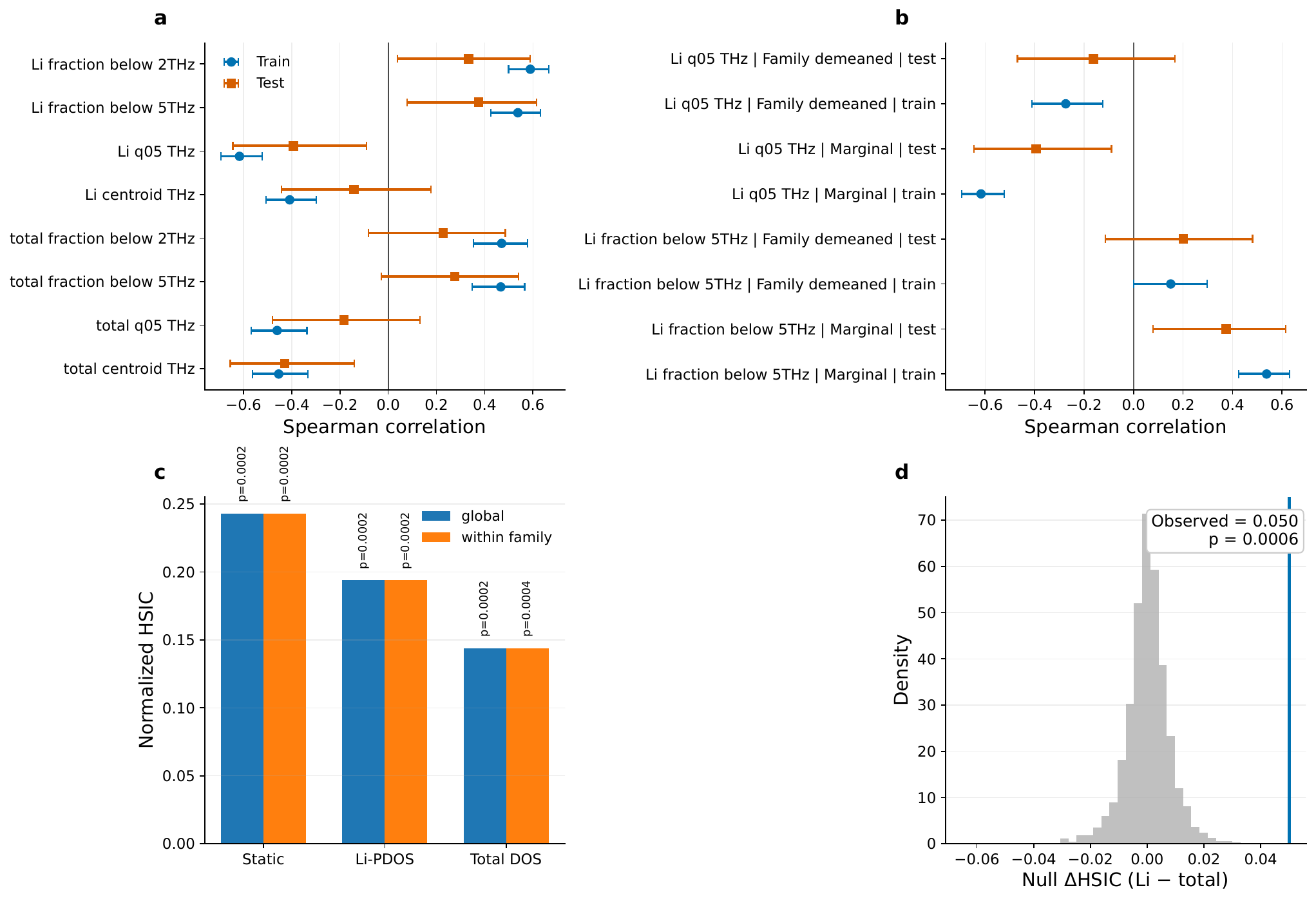}
\caption{Scalar and whole-spectrum dependence. (a) Train/test replication of selected Li and total-DOS descriptors. (b) Marginal, family-demeaned, and static-residual descriptor associations. (c) Normalized HSIC for static, Li-Wasserstein, and total-Wasserstein representations under global, within-family, and residual tests. (d) Paired Li-minus-total HSIC result for the primary cohort relative to its permutation null. Complete primary/strict/exact data and HSIC sensitivity are shown in Figures S26-S29.}
\label{fig:dependence}
\end{figure*}

The pooled descriptor effects are not equivalent to a universal within-family law. In the primary training cohort, the marginal correlation for Li weight below 5 THz is $\rho=0.538$, but it decreases to approximately 0.150 after family demeaning; the corresponding test interval crosses zero (Figure~\ref{fig:dependence}). The Li $q_{05}$ association also weakens after family adjustment. This attenuation is expected because both conductivity and spectral shape are strongly organized by chemistry. A descriptor can therefore be useful for cross-material screening while carrying less information for substitutions within a narrowly defined family.

Whole-spectrum HSIC confirms significant nonlinear dependence under both global and within-family permutations (Figure~\ref{fig:dependence}c). In the primary cohort, normalized HSIC is 0.194 for Li-Wasserstein and 0.144 for total-Wasserstein; their paired difference is 0.0501 with one-sided permutation $p=6\times10^{-4}$. The strict cohort gives a difference of 0.0395 ($p=1.2\times10^{-3}$). However, the exact cohort reverses the ordering: Li HSIC is 0.125 and total HSIC is 0.184. The smaller exact set has a different family composition, so this reversal prevents a universal claim that Li projection always has greater dependence than total DOS.

Residual HSIC provides a second nuance. In the strict and exact cohorts, Li-Wasserstein remains associated with out-of-fold static residuals under within-family permutations ($p=0.0052$ and 0.0058), whereas the primary-cohort result is weaker ($p=0.069$). Taken together, the evidence supports complementary information in the Li distribution but shows that the magnitude and even the Li-versus-total ordering depend on cohort composition. 

\subsection{Wasserstein \LiPDOS Gives the Most Stable Held-Out Prediction}
\begin{figure*}[t]
\centering
\includegraphics[width=\textwidth]{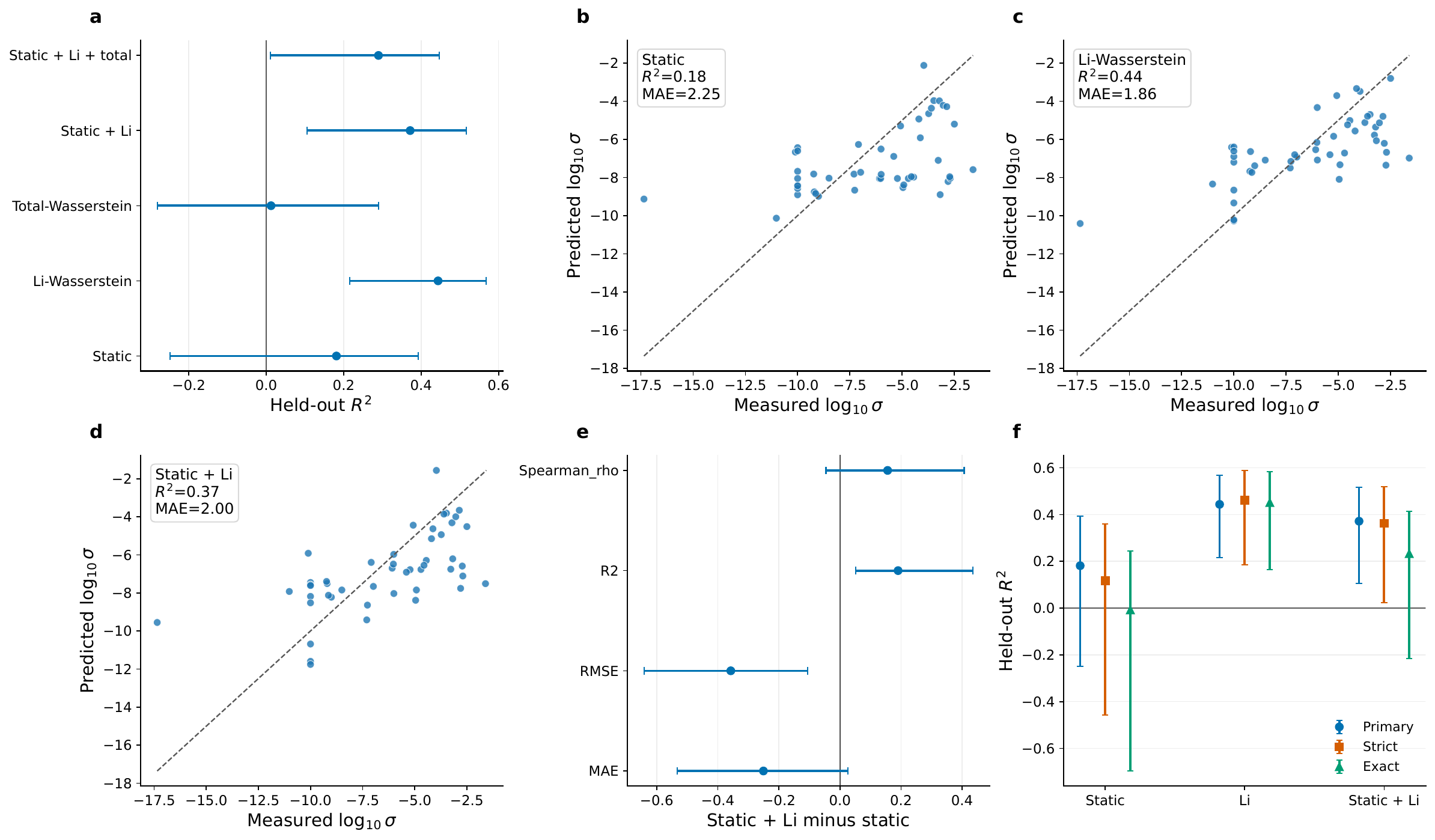}
\caption{Held-out prediction and audit-cohort sensitivity. (a) Official-test MAE for static, Li-Wasserstein, total-Wasserstein, and static+Li models. (b--d) Observed versus predicted log conductivity for the primary test cohort using the static, Li-Wasserstein, and static+Li models. (e) Paired bootstrap differences relative to the static model. (f) $R^2$ across primary, strict, and exact audit cohorts. Hyperparameters for strict and exact sensitivity models were fixed from primary-training selection.}
\label{fig:prediction}
\end{figure*}

The official-test comparison in Figure~\ref{fig:prediction} shows a clear separation among representations. In the primary cohort, the static RBF baseline gives mean absolute error (MAE) 2.250, rooot-mean squared error (RMSE) 2.883, $R^2=0.181$, and Spearman $\rho=0.476$. Total-DOS Wasserstein gives MAE 2.344 and $R^2=0.012$. The standalone Li-PDOS Wasserstein model improves to MAE 1.859, RMSE 2.376, $R^2=0.444$, and $\rho=0.620$. Static+Li fusion gives $R^2=0.372$ and $\rho=0.632$ but does not improve MAE or $R^2$ over Li alone. The principal empirical result is therefore not synergy; it is that the distribution-aware mobile-ion representation contains information inadequately represented by the tested static and total-DOS baselines.

The Li model is unusually stable to the audit definition. Its test $R^2$ is 0.444 in the primary cohort, 0.462 in the strict cohort, and 0.451 in the exact cohort, with MAE near 1.85 in all three. By contrast, the static $R^2$ declines from 0.181 to 0.116 and then $-0.008$. In the strict cohort, Li versus static improves MAE by $-0.589$ (95\% CI $-1.011$ to $-0.149$), RMSE by $-0.667$, and $R^2$ by 0.346. In the exact cohort, the differences are $-0.905$, $-0.864$, and 0.459, respectively. The corresponding total-DOS differences from static are not statistically resolved.

An MAE near 1.86 log units is too large for precise material-level conductivity prediction and reflects experimental heterogeneity, imperfect structural correspondence, temperature and processing variation, censoring, and the limitations of a harmonic MLIP spectrum. The model is better interpreted as a comparative ranking and representation test that can prioritize candidates before more expensive MLIP molecular dynamics (MLIP-MD) or first-principles diffusion calculations. In that hierarchy, static descriptors are inexpensive but dynamically incomplete; harmonic Li-PDOS adds mobile-ion-resolved information at moderate cost; finite-temperature MLIP-MD or ab-initio MD (AIMD) is required to estimate diffusion and anharmonic transport more directly.\cite{sendek2017,sendek2019,jalem2018,muy2019,kim2024,aghoghovbia2026,du2025}

\subsection{Implications for Solid-Electrolyte Screening}
For an applied energy-materials workflow, the practical question is not whether a single descriptor is universally mechanistic, but whether it adds reliable information at a useful computational cost. The present results suggest three design principles. First, mobile-ion projections should be retained when phonons are used for screening; a total DOS can mask Li-specific changes even when it remains informative about framework softness. Second, the spectrum should be treated as an ordered distribution rather than reduced immediately to a band center. Scalar features reproduce, but the full Li-Wasserstein model is more stable across audit cohorts. Third, structural provenance must accompany spectral prediction. Disorder integerization, vacancy preservation, exact-phase matching, and family composition materially affect inferential conclusions.

The framework is most naturally used as an intermediate screen. A low-frequency Li redistribution or favorable Wasserstein similarity to known conductors can prioritize structures for temperature-dependent MLIP-MD, enhanced sampling, or AIMD. Conversely, a high model score should not override instability, unfavorable electrochemical compatibility, or a failed composition audit. The audit and reference comparisons also provide a template for deploying universal MLIPs in other energy-material problems: verify the input structure, compare species-resolved observables rather than only aggregate quantities, and carry model uncertainty into the claim.

\section{Conclusions}
An audit-filtered MatterSim--Phonopy analysis of the OBELiX experimental dataset shows that lithium-projected phonon distributions contain reproducible information about room-temperature ionic conductivity. Conductive materials exhibit increased normalized Li weight at low frequency and earlier cumulative spectral onset, with directional replication in the official test split. A Wasserstein kernel on the complete Li-PDOS substantially outperforms total DOS and a static composition--structure baseline and retains nearly unchanged performance across primary, strict, and exact composition cohorts.

The boundaries of the result are equally important. Independent DFT references reveal systematic, chemistry-dependent MLIP softening that is larger for Li projections than for total DOS. Family demeaning attenuates scalar effects, and the Li-versus-total HSIC ordering reverses in the exact cohort. Harmonic projected spectra therefore provide a robust comparative descriptor, not proof that low-frequency phonons universally cause fast transport. Future work should connect the identified spectral regions to finite-temperature diffusion, mode-resolved hopping coordinates, and anharmonic Li--framework coupling using MLIP-MD and targeted first-principles validation.

\section{Supporting Information}
The Supporting Information is available free of charge and contains audit-cohort sensitivities; frequency-bin-width, censoring, nonlinearity, family-transfer, scalar-versus-distribution, frequency-warp, imaginary-mode, fusion, kernel-weight, and residual analyses; all external reference overlays; convergence calculations; family-specific associations; censored-regression sensitivity; static-residual frequency associations; spectral quality control; and the full model comparison. Exact plotted-data CSV files accompany every figure in the public repository.

\section{Data Availability Statement}
The analysis code, manifests, audit workflow, and machine-readable result tables are available at \url{https://github.com/Gajendra9843/MLIP}. The OBELiX source data are available through the repository reported by Therrien et al.\cite{therrien2026}.

\section{Author Contributions}
G.B. performed the computational calculations and data analysis. R.P. conceived and supervised the study. Both authors interpreted the results and wrote the manuscript. 

\section{Competing Interests}
The authors declare no competing financial interest.

\bibliography{references}

@article{manthiram2017,
  author = {Manthiram, Arumugam and Yu, Xingwen and Wang, Shaofei},
  title = {Lithium Battery Chemistries Enabled by Solid-State Electrolytes},
  journal = {Nat. Rev. Mater.},
  year = {2017}, volume = {2}, pages = {16103},
  doi = {10.1038/natrevmats.2016.103}
}

@article{janek2016,
  author = {Janek, Jürgen and Zeier, Wolfgang G.},
  title = {A Solid Future for Battery Development},
  journal = {Nat. Energy}, year = {2016}, volume = {1}, pages = {16141},
  doi = {10.1038/nenergy.2016.141}
}

@article{bachman2016,
  author = {Bachman, John C. and Muy, Sokseiha and Grimaud, Alexis and Chang, Hao-Hsun and Pour, Nima and Lux, Simon F. and Paschos, Odysseas and Maglia, Filippo and Lupart, Saskia and Lamp, Peter and Giordano, Livia and Shao-Horn, Yang},
  title = {Inorganic Solid-State Electrolytes for Lithium Batteries: Mechanisms and Properties Governing Ion Conduction},
  journal = {Chem. Rev.}, year = {2016}, volume = {116}, pages = {140--162},
  doi = {10.1021/acs.chemrev.5b00563}
}

@article{famprikis2019,
  author = {Famprikis, Theodosios and Canepa, Pieremanuele and Dawson, James A. and Islam, M. Saiful and Masquelier, Christian},
  title = {Fundamentals of Inorganic Solid-State Electrolytes for Batteries},
  journal = {Nat. Mater.}, year = {2019}, volume = {18}, pages = {1278--1291},
  doi = {10.1038/s41563-019-0431-3}
}

@article{albertus2018,
  author = {Albertus, Paul and Babinec, Susan and Litzelman, Scott and Newman, Aron},
  title = {Status and Challenges in Enabling the Lithium Metal Electrode for High-Energy and Low-Cost Rechargeable Batteries},
  journal = {Nat. Energy}, year = {2018}, volume = {3}, pages = {16--21},
  doi = {10.1038/s41560-017-0047-2}
}

@article{randau2020,
  author = {Randau, Simon and Weber, David A. and Kötz, Otto and Koerver, Raimund and Braun, Patrick and Weber, André and Ivers-Tiffée, Ellen and Adermann, Tobias and Kulisch, Jörg and Zeier, Wolfgang G. and Richter, Franz H. and Janek, Jürgen},
  title = {Benchmarking the Performance of All-Solid-State Lithium Batteries},
  journal = {Nat. Energy}, year = {2020}, volume = {5}, pages = {259--270},
  doi = {10.1038/s41560-020-0565-1}
}

@article{banerjee2020,
  author = {Banerjee, Abhik and Wang, Xuefeng and Fang, Chengcheng and Wu, Erik A. and Meng, Ying Shirley},
  title = {Interfaces and Interphases in All-Solid-State Batteries with Inorganic Solid Electrolytes},
  journal = {Chem. Rev.}, year = {2020}, volume = {120}, pages = {6878--6933},
  doi = {10.1021/acs.chemrev.0c00101}
}

@article{xiao2020,
  author = {Xiao, Yanhao and Wang, Yan and Bo, Shou-Hang and Kim, Jae Chul and Miara, Lincoln J. and Ceder, Gerbrand},
  title = {Understanding Interface Stability in Solid-State Batteries},
  journal = {Nat. Rev. Mater.}, year = {2020}, volume = {5}, pages = {105--126},
  doi = {10.1038/s41578-019-0157-5}
}

@article{zhao2020,
  author = {Zhao, Qian and Stalin, Swetha and Zhao, Cheng-Zheng and Archer, Lynden A.},
  title = {Designing Solid-State Electrolytes for Safe, Energy-Dense Batteries},
  journal = {Nat. Rev. Mater.}, year = {2020}, volume = {5}, pages = {229--252},
  doi = {10.1038/s41578-019-0165-5}
}

@article{kerman2017,
  author = {Kerman, Kian and Luntz, Alan and Viswanathan, Viswanathan and Chiang, Yet-Ming and Chen, Zhongwei},
  title = {Review---Practical Challenges Hindering the Development of Solid State Li Ion Batteries},
  journal = {J. Electrochem. Soc.}, year = {2017}, volume = {164}, pages = {A1731--A1744},
  doi = {10.1149/2.1571707jes}
}

@article{porz2017,
  author = {Porz, Lukas and Swamy, Tirumala and Sheldon, Brian W. and Rettenwander, Daniel and Frömling, Till and Thaman, Henry L. and Berendts, Stefan and Uecker, Reinhard and Carter, W. Craig and Chiang, Yet-Ming},
  title = {Mechanism of Lithium Metal Penetration through Inorganic Solid Electrolytes},
  journal = {Adv. Energy Mater.}, year = {2017}, volume = {7}, pages = {1701003},
  doi = {10.1002/aenm.201701003}
}

@article{han2019,
  author = {Han, Fudong and Westover, Andrew S. and Yue, Jia and Fan, Xueliang and Wang, Fei and Chi, Ming and Leonard, Dean N. and Dudney, Nancy J. and Wang, Howard and Wang, Chunsheng},
  title = {High Electronic Conductivity as the Origin of Lithium Dendrite Formation within Solid Electrolytes},
  journal = {Nat. Energy}, year = {2019}, volume = {4}, pages = {187--196},
  doi = {10.1038/s41560-018-0312-z}
}

@article{wang2015,
  author = {Wang, Yan and Richards, William Davidson and Ong, Shyue Ping and Miara, Lincoln J. and Kim, Jae Chul and Mo, Yifei and Ceder, Gerbrand},
  title = {Design Principles for Solid-State Lithium Superionic Conductors},
  journal = {Nat. Mater.}, year = {2015}, volume = {14}, pages = {1026--1031},
  doi = {10.1038/nmat4369}
}

@article{muy2018,
  author = {Muy, Sokseiha and Bachman, John C. and Giordano, Livia and Chang, Hao-Hsun and Abernathy, Douglas L. and Bansal, Dipanshu and Delaire, Olivier and Hori, Satoshi and Kanno, Ryoji and Maglia, Filippo and Lupart, Saskia and Lamp, Peter and Shao-Horn, Yang},
  title = {Tuning Mobility and Stability of Lithium Ion Conductors Based on Lattice Dynamics},
  journal = {Energy Environ. Sci.}, year = {2018}, volume = {11}, pages = {850--859},
  doi = {10.1039/C7EE03364H}
}

@article{krauskopf2017,
  author = {Krauskopf, Thorben and Pompe, Constantin and Kraft, Marvin A. and Zeier, Wolfgang G.},
  title = {Influence of Lattice Dynamics on Na+ Transport in Na3PS4-xSex},
  journal = {Chem. Mater.}, year = {2017}, volume = {29}, pages = {8859--8869},
  doi = {10.1021/acs.chemmater.7b03474}
}

@article{krauskopf2018,
  author = {Krauskopf, Thorben and Muy, Sokseiha and Culver, Sean P. and Ohno, Saneyuki and Delaire, Olivier and Shao-Horn, Yang and Zeier, Wolfgang G.},
  title = {Comparing the Descriptors for Investigating the Influence of Lattice Dynamics on Ionic Transport Using the Superionic Conductor Na3PS4-xSex},
  journal = {J. Am. Chem. Soc.}, year = {2018}, volume = {140}, pages = {14464--14473},
  doi = {10.1021/jacs.8b09340}
}

@article{muy2019,
  author = {Muy, Sokseiha and Voss, Johannes and Schlem, Roman and Koerver, Raimund and Sedlmaier, Stefan J. and Maglia, Filippo and Lamp, Peter and Zeier, Wolfgang G. and Shao-Horn, Yang},
  title = {High-Throughput Screening of Solid-State Li-Ion Conductors Using Lattice-Dynamics Descriptors},
  journal = {iScience}, year = {2019}, volume = {16}, pages = {270--282},
  doi = {10.1016/j.isci.2019.05.036}
}

@article{muy2021,
  author = {Muy, Sokseiha and Schlem, Roman and Shao-Horn, Yang and Zeier, Wolfgang G.},
  title = {Phonon--Ion Interactions: Designing Ion Mobility Based on Lattice Dynamics},
  journal = {Adv. Energy Mater.}, year = {2021}, volume = {11}, pages = {2002787},
  doi = {10.1002/aenm.202002787}
}

@article{sagotra2019,
  author = {Sagotra, Arun K. and Chu, Dewei and Cazorla, Claudio},
  title = {Influence of Lattice Dynamics on Lithium-Ion Conductivity: A First-Principles Study},
  journal = {Phys. Rev. Mater.}, year = {2019}, volume = {3}, pages = {035405},
  doi = {10.1103/PhysRevMaterials.3.035405}
}

@article{gordiz2021,
  author = {Gordiz, Kiarash and Muy, Sokseiha and Zeier, Wolfgang G. and Shao-Horn, Yang and Henry, Asegun},
  title = {Enhancement of Ion Diffusion by Targeted Phonon Excitation},
  journal = {Cell Rep. Phys. Sci.}, year = {2021}, volume = {2}, pages = {100431},
  doi = {10.1016/j.xcrp.2021.100431}
}

@article{xu2022,
  author = {Xu, Zhenming and Chen, Xi and Zhu, Hong and Li, Xin},
  title = {Anharmonic Cation--Anion Coupling Dynamics Assisted Lithium-Ion Diffusion in Sulfide Solid Electrolytes},
  journal = {Adv. Mater.}, year = {2022}, volume = {34}, pages = {2207411},
  doi = {10.1002/adma.202207411}
}

@article{song2024,
  author = {Song, Tao and Lin, Yuxiao and Wang, Da and Chen, Qianli and Ling, Chen and Shi, Siqi},
  title = {Renewing Fundamental Understanding of Ionic Transport in Inorganic Crystalline Solid-State Electrolytes from the Perspective of Lattice Dynamics},
  journal = {Adv. Energy Mater.}, year = {2024}, volume = {14}, pages = {2302440},
  doi = {10.1002/aenm.202302440}
}

@article{ouyang2024,
  author = {Ouyang, Runxin and Yang, Yu and Guan, Chaohong and Zhu, Hong},
  title = {Phonon--Lithium Ion Interactions: A Case Study of LiM(SeO3)2 (M = Al, Ga, In, Sc, Y, and La)},
  journal = {ACS Appl. Mater. Interfaces}, year = {2024}, volume = {16}, pages = {55240--55247},
  doi = {10.1021/acsami.4c09985}
}

@article{ding2025,
  author = {Ding, Jingxuan and Gupta, Mayanak K. and Rosenbach, Carolin and Lin, Hung-Min and Osti, Naresh C. and Abernathy, Douglas L. and Zeier, Wolfgang G. and Delaire, Olivier},
  title = {Liquid-like Dynamics in a Solid-State Lithium Electrolyte},
  journal = {Nat. Phys.}, year = {2025}, volume = {21}, pages = {118--125},
  doi = {10.1038/s41567-024-02707-6}
}

@article{pham2026,
  author = {Pham, Kim H. and Gordiz, Kiarash and Spear, Nathaniel A. and Lin, Andrew K. and Michelsen, Jonathan M. and Liu, Hanzhe and Vivona, Daniele and Blake, Geoffrey A. and Shao-Horn, Yang and Henry, Asegun and See, Kimberly A. and Cushing, Scott K.},
  title = {Correlated Terahertz Phonon--Ion Interactions Control Ion Conduction in a Solid Electrolyte},
  journal = {Mater. Horiz.}, year = {2026}, volume = {13}, pages = {3355--3375},
  doi = {10.1039/D5MH01990G}
}

@article{li2015,
  author = {Li, Xinyu and Benedek, Nicole A.},
  title = {Enhancement of Ionic Transport in Complex Oxides through Soft Lattice Modes and Epitaxial Strain},
  journal = {Chem. Mater.}, year = {2015}, volume = {27}, pages = {2647--2652},
  doi = {10.1021/acs.chemmater.5b00445}
}

@article{greene2024,
  author = {Greene, Samuel M. and Siegel, Donald J.},
  title = {Assessing Correlations between Phonon Features and Cation Migration Barriers in Multivalent Solid Electrolytes},
  journal = {Chem. Mater.}, year = {2024}, volume = {36}, pages = {7476--7486},
  doi = {10.1021/acs.chemmater.4c01468}
}

@article{kim2024,
  author = {Kim, Jiyeon and Lee, Donggeon and Lee, Dongwoo and Li, Xin and Lee, Yea-Lee and Kim, Sooran},
  title = {Machine Learning Prediction Models for Solid Electrolytes Based on Lattice Dynamics Properties},
  journal = {J. Phys. Chem. Lett.}, year = {2024}, volume = {15}, pages = {5914--5922},
  doi = {10.1021/acs.jpclett.4c00995}
}

@article{jaafreh2024,
  author = {Jaafreh, Russlan and Pereznieto, Santiago and Jeong, Seonghun and Widiantara, I. Putu and Oh, Jeong Moo and Kang, Jee-Hyun and Mun, Junyoung and Ko, Young Gun and Kim, Jung-Gu and Hamad, Kotiba},
  title = {Phonon DOS-Based Machine Learning Model for Designing High-Performance Solid Electrolytes in Li-Ion Batteries},
  journal = {Int. J. Energy Res.}, year = {2024}, volume = {2024}, pages = {2138847},
  doi = {10.1155/2024/2138847}
}

@article{therrien2026,
  author = {Therrien, Félix and Abou Haibeh, Jamal and Sharma, Divya and Hendley, Rhiannon and Mungai, Leah Wairimu and Sun, Sun and Tchagang, Alain and Su, Jiang and Huberman, Samuel and Bengio, Yoshua and Guo, Hongyu and Hernández-García, Alex and Shin, Homin},
  title = {OBELiX: A Curated Dataset of Crystal Structures and Experimentally Measured Ionic Conductivities for Lithium Solid-State Electrolytes},
  journal = {Digital Discovery}, year = {2026}, volume = {5}, pages = {910--918},
  doi = {10.1039/D5DD00441A}
}

@article{sendek2017,
  author = {Sendek, Austin D. and Yang, Qing and Cubuk, Ekin D. and Duerloo, Karel-Alexander N. and Cui, Yi and Reed, Evan J.},
  title = {Holistic Computational Structure Screening of More than 12,000 Candidates for Solid Lithium-Ion Conductor Materials},
  journal = {Energy Environ. Sci.}, year = {2017}, volume = {10}, pages = {306--320},
  doi = {10.1039/C6EE02697D}
}

@article{jalem2018,
  author = {Jalem, Randy and Kanamori, Katsumi and Takeuchi, Ichiro and Nakayama, Masanobu and Yamasaki, Hiroyuki and Saito, Takuya},
  title = {Bayesian-Driven First-Principles Calculations for Accelerating Exploration of Fast Ion Conductors for Rechargeable Battery Application},
  journal = {Sci. Rep.}, year = {2018}, volume = {8}, pages = {5845},
  doi = {10.1038/s41598-018-23827-3}
}

@article{he2020,
  author = {He, Xingfeng and Zhu, Yizhou and Mo, Yifei},
  title = {Origin of Fast Ion Diffusion in Super-Ionic Conductors},
  journal = {Nat. Commun.}, year = {2017}, volume = {8}, pages = {15893},
  doi = {10.1038/ncomms15893}
}

@article{yang2024,
  author = {Yang, Han and Hu, Chenxi and Zhou, Yichi and Liu, Xixian and Shi, Yu and Li, Jielan and Li, Guanzhi and Chen, Zekun and Chen, Shuizhou and Zeni, Claudio and Horton, Matthew and Pinsler, Robert and Fowler, Andrew and Zügner, Daniel and Xie, Tian and Smith, Jake and Sun, Lixin and Wang, Qian and Kong, Lingyu and Liu, Chang and Hao, Hongxia and Lu, Ziheng},
  title = {MatterSim: A Deep Learning Atomistic Model Across Elements, Temperatures and Pressures},
  journal = {arXiv}, year = {2024}, pages = {2405.04967},
  doi = {10.48550/arXiv.2405.04967}
}

@article{loew2025,
  author = {Loew, Antoine and Sun, Dewen and Wang, Hai-Chen and Botti, Silvana and Marques, Miguel A. L.},
  title = {Universal Machine Learning Interatomic Potentials Are Ready for Phonons},
  journal = {npj Comput. Mater.}, year = {2025}, volume = {11}, pages = {178},
  doi = {10.1038/s41524-025-01650-1}
}

@article{deng2025,
  author = {Deng, Bowen and Zhong, Peichen and Jun, KyuJung and Riebesell, Janosh and Han, Kevin and Bartel, Christopher J. and Ceder, Gerbrand},
  title = {Systematic Softening in Universal Machine Learning Interatomic Potentials},
  journal = {npj Comput. Mater.}, year = {2025}, volume = {11}, pages = {9},
  doi = {10.1038/s41524-024-01500-6}
}

@article{du2025,
  author = {Du, Hongwei and Huang, Xiang and Hui, Jian and Zhang, Lanting and Zhou, Yuanxun and Wang, Hong},
  title = {Assessment and Application of Universal Machine Learning Interatomic Potentials in Solid-State Electrolyte Research},
  journal = {ACS Mater. Lett.}, year = {2025}, volume = {7}, pages = {3403--3412},
  doi = {10.1021/acsmaterialslett.5c00336}
}

@article{parlinski1997,
  author = {Parlinski, Krzysztof and Li, Z. Q. and Kawazoe, Yoshiyuki},
  title = {First-Principles Determination of the Soft Mode in Cubic ZrO2},
  journal = {Phys. Rev. Lett.}, year = {1997}, volume = {78}, pages = {4063--4066},
  doi = {10.1103/PhysRevLett.78.4063}
}

@article{togo2023jpsj,
  author = {Togo, Atsushi},
  title = {First-Principles Phonon Calculations with Phonopy and Phono3py},
  journal = {J. Phys. Soc. Jpn.}, year = {2023}, volume = {92}, pages = {012001},
  doi = {10.7566/JPSJ.92.012001}
}

@article{togo2023jpcm,
  author = {Togo, Atsushi and Chaput, Laurent and Tadano, Terumasa and Tanaka, Isao},
  title = {Implementation Strategies in Phonopy and Phono3py},
  journal = {J. Phys.: Condens. Matter}, year = {2023}, volume = {35}, pages = {353001},
  doi = {10.1088/1361-648X/acd831}
}

@article{jain2013,
  author = {Jain, Anubhav and Ong, Shyue Ping and Hautier, Geoffroy and Chen, Wei and Richards, William Davidson and Dacek, Stephen and Cholia, Shreyas and Gunter, Dan and Skinner, David and Ceder, Gerbrand and Persson, Kristin A.},
  title = {Commentary: The Materials Project: A Materials Genome Approach to Accelerating Materials Innovation},
  journal = {APL Mater.}, year = {2013}, volume = {1}, pages = {011002},
  doi = {10.1063/1.4812323}
}

@article{ong2013,
  author = {Ong, Shyue Ping and Richards, William Davidson and Jain, Anubhav and Hautier, Geoffroy and Kocher, Michael and Cholia, Shreyas and Gunter, Dan and Chevrier, Vincent L. and Persson, Kristin A. and Ceder, Gerbrand},
  title = {Python Materials Genomics (pymatgen): A Robust, Open-Source Python Library for Materials Analysis},
  journal = {Comput. Mater. Sci.}, year = {2013}, volume = {68}, pages = {314--319},
  doi = {10.1016/j.commatsci.2012.10.028}
}

@article{larsen2017,
  author = {Larsen, Ask Hjorth and Mortensen, Jens Jørgen and Blomqvist, Jakob and Castelli, Ivano E. and Christensen, Rune and Dułak, Marcin and Friis, Jesper and Groves, Michael N. and Hammer, Bjørk and Hargus, Cory and Hermes, Eric D. and Jennings, Paul C. and Jensen, Peter Bjerre and Kermode, James and Kitchin, John R. and Kolsbjerg, Esben L. and Kubal, Joseph and Kaasbjerg, Kristen and Lysgaard, Steen and Maronsson, Jón Bergmann and Maxson, Tristan and Olsen, Thomas and Pastewka, Lars and Peterson, Andrew and Rostgaard, Carsten and Schiøtz, Jakob and Schütt, Ole and Strange, Mikkel and Thygesen, Kristian S. and Vegge, Tejs and Vilhelmsen, Lasse and Walter, Michael and Zeng, Zhenhua and Jacobsen, Karsten W.},
  title = {The Atomic Simulation Environment---A Python Library for Working with Atoms},
  journal = {J. Phys.: Condens. Matter}, year = {2017}, volume = {29}, pages = {273002},
  doi = {10.1088/1361-648X/aa680e}
}

@article{aitchison1982,
  author = {Aitchison, John},
  title = {The Statistical Analysis of Compositional Data},
  journal = {J. R. Stat. Soc. B}, year = {1982}, volume = {44}, pages = {139--160},
  doi = {10.1111/j.2517-6161.1982.tb01195.x}
}

@inproceedings{kolouri2016,
  author = {Kolouri, Soheil and Zou, Yang and Rohde, Gustavo K.},
  title = {Sliced Wasserstein Kernels for Probability Distributions},
  booktitle = {Proc. IEEE Conf. Comput. Vis. Pattern Recognit.}, year = {2016}, pages = {5258--5267},
  doi = {10.1109/CVPR.2016.568}
}

@article{bachoc2018,
  author = {Bachoc, François and Gamboa, Fabrice and Loubes, Jean-Michel and Venet, Nil},
  title = {A Gaussian Process Regression Model for Distribution Inputs},
  journal = {IEEE Trans. Inf. Theory}, year = {2018}, volume = {64}, pages = {6620--6637},
  doi = {10.1109/TIT.2017.2762322}
}

@incollection{gretton2005,
  author = {Gretton, Arthur and Bousquet, Olivier and Smola, Alexander J. and Schölkopf, Bernhard},
  title = {Measuring Statistical Dependence with Hilbert-Schmidt Norms},
  booktitle = {Algorithmic Learning Theory}, year = {2005}, volume = {3734}, pages = {63--77},
  doi = {10.1007/11564089_7}
}

@article{efron1979,
  author = {Efron, Bradley}, title = {Bootstrap Methods: Another Look at the Jackknife},
  journal = {Ann. Stat.}, year = {1979}, volume = {7}, pages = {1--26},
  doi = {10.1214/aos/1176344552}
}

@article{benjamini1995,
  author = {Benjamini, Yoav and Hochberg, Yosef},
  title = {Controlling the False Discovery Rate: A Practical and Powerful Approach to Multiple Testing},
  journal = {J. R. Stat. Soc. B}, year = {1995}, volume = {57}, pages = {289--300},
  doi = {10.1111/j.2517-6161.1995.tb02031.x}
}

@article{kraskov2004,
  author = {Kraskov, Alexander and Stögbauer, Harald and Grassberger, Peter},
  title = {Estimating Mutual Information},
  journal = {Phys. Rev. E}, year = {2004}, volume = {69}, pages = {066138},
  doi = {10.1103/PhysRevE.69.066138}
}

@article{szekely2007,
  author = {Székely, Gábor J. and Rizzo, Maria L. and Bakirov, Nail K.},
  title = {Measuring and Testing Dependence by Correlation of Distances},
  journal = {Ann. Stat.}, year = {2007}, volume = {35}, pages = {2769--2794},
  doi = {10.1214/009053607000000505}
}

@inproceedings{chen2016,
  author = {Chen, Tianqi and Guestrin, Carlos}, title = {XGBoost: A Scalable Tree Boosting System},
  booktitle = {Proc. 22nd ACM SIGKDD Int. Conf. Knowl. Discovery Data Mining}, year = {2016}, pages = {785--794},
  doi = {10.1145/2939672.2939785}
}

@article{hoerl1970,
  author = {Hoerl, Arthur E. and Kennard, Robert W.}, title = {Ridge Regression: Biased Estimation for Nonorthogonal Problems},
  journal = {Technometrics}, year = {1970}, volume = {12}, pages = {55--67},
  doi = {10.1080/00401706.1970.10488634}
}

@article{lu2022,
  author = {Lu, Zhuole and Adeli, Parvin and Yim, Chae-Ho and Mercier, Patrick and Abu-Lebdeh, Yaser and Singh, Chandra Veer},
  title = {Automatically Capturing Key Features for Predicting Superionic Conductivity of Solid-State Electrolytes Using a Neural Network},
  journal = {ACS Appl. Energy Mater.}, year = {2022}, volume = {5}, pages = {8042--8048},
  doi = {10.1021/acsaem.2c00493}
}

@article{sendek2019,
  author = {Sendek, Austin D. and Cubuk, Ekin D. and Antoniuk, Evan R. and Cheon, Gowoon and Cui, Yi and Reed, Evan J.},
  title = {Machine Learning-Assisted Discovery of Solid Li-Ion Conducting Materials},
  journal = {Chem. Mater.}, year = {2019}, volume = {31}, pages = {342--352},
  doi = {10.1021/acs.chemmater.8b03272}
}

@article{aghoghovbia2026,
  author = {Aghoghovbia, Ogheneyoma and Rurali, Riccardo and Al-Fahdi, Mohammed and Hu, Ming},
  title = {Unlocking Lithium Superionic Conduction via Phonon Softness Descriptors: A High-Throughput Machine Learning Paradigm},
  journal = {Chem. Mater.}, year = {2026}, volume = {38}, pages = {1689--1705},
  doi = {10.1021/acs.chemmater.5c02214}
}

@misc{togoNIMSPhononDB,
  author       = {Togo, Atsushi},
  title        = {{MDR Phonon Calculation Database}},
  year         = {2025},
  publisher    = {National Institute for Materials Science},
  howpublished = {NIMS Materials Data Repository},
  doi          = {10.48505/nims.4197},
  note         = {Harmonic phonon calculation database generated using VASP
                  and phonopy; accessed July 19, 2026}
}

@article{togo2023phonopy,
  author  = {Togo, Atsushi and Chaput, Laurent and Tadano, Terumasa
             and Tanaka, Isao},
  title   = {Implementation Strategies in Phonopy and Phono3py},
  journal = {Journal of Physics: Condensed Matter},
  year    = {2023},
  volume  = {35},
  number  = {35},
  pages   = {353001},
  doi     = {10.1088/1361-648X/acd831}
}
\end{document}